\documentclass[12pt,english]{article}
\usepackage{amssymb}

\makeatletter

\makeatletter

\makeatletter

\AtBeginDocument{\DeclareFontEncoding{T2A}{}{}}

\makeatletter

\AtBeginDocument{\DeclareFontEncoding{T2A}{}{}}

\makeatletter

\AtBeginDocument{\DeclareFontEncoding{T2A}{}{}}

\makeatletter

\makeatletter

\makeatletter

\newtheorem{theorem}{Theorem}

\newtheorem{lemma}{Lemma}

\date{}

\begin{document}

\title{Fixed points for one-dimensional particle system with strong interaction}

\author{V. A. Malyshev}
\maketitle
\begin{abstract}
We consider hamiltonian $N$ particle system on the finite segment
with nearest-neighbor Coulomb interaction and external force $F$.
We study the fixed points of such system and show that the distances
between neighbors are asymptotically, for large $N$, the same for
any $F$. 
\end{abstract}

\section{Introduction}

The main goal of the paper is to present a peculiar equilibrium model
of large system of interacting particles with strong interaction.
We study properties of this system for zero temperature, that is we
study ground states of the model. Moreover, it exhibits phenomena
somewhat unusual for the modern mathematical statistical physics.
Namely, the local equilibrium (defining the global equilibrium) in
this model lives on the scale much smaller than normal microscale,
see more discussion at the final remarks.

We consider the system of $N$ classical identical point particles
on the interval $[0,L]\in R$\[
0\leq x_{1}(t)<...<x_{N}(t)\leq L\]
 The dynamics is defined by the system of $N$ equations \begin{equation}
m\frac{d^{2}x_{i}}{dt^{2}}=-\frac{\partial U}{\partial x_{i}}+F(x_{i})-A\frac{dx_{i}}{dt}\label{main_eq}\end{equation}
 where the interaction is

\[
U(x_{1},...,x_{N})=\sum_{i=2}^{N}V(x_{i}-x_{i-1})\]
 It is assumed that the potential $V(x)=V(-x)>0$ is repulsive, that
is \[
f(r)=-\frac{dV(r)}{dr}>0,r=|x|\]
It is also assumed that $f$ is sufficiently smooth, monotone decreasing
and\begin{equation}
f(r)\sim_{r\to0}\alpha r^{-a},\alpha>0,a>1\label{asympt_f}\end{equation}
The example is\begin{equation}
f(r)=\alpha r^{-a}\label{powerLaw}\end{equation}
Further on we take $\alpha=1$. 

It is also assumed that the external force $F$ is continuous function
on $[0,L]$. We assume completely inelastic boundary conditions, that
is when a particle $x_{1}(t)$ reaches point $0$ having the velocity
$v_{1}(t)<0$ then immediately its velocity $v_{1}(t+0)$ becomes
$0$, the particle stays at $0$ until the force $F(0)-f(x_{2}(t))$
becomes positive. Similarly for the particle $x_{N}(t)$ at the point
$L$.

It follows from the elementary theory of ODE that for any initial
data\[
(x_{i}(0),v_{i}(0)),i=1,...,N,x_{i}(0)<x_{i+1}(0)\]
 the solution exists and is unique for $t\in[0,\infty)$. Moreover,
the order of particles cannot change. We will study the fixed points
of this system.

For zero external force $F(x)\equiv0$ the point $(x_{1},...,x_{N})$
with $x_{k}=\frac{L(k-1)}{N-1}$ is evidently the unique fixed point.
In fact, the particles $x_{1},x_{N}$ reaching the boundary stay there
forever. Moreover, the force acting on particle $i=2,...,N-1$ is
zero only if $x_{i}-x_{i-1}=x_{i+1}-x_{i}$. This is only possible
if\begin{equation}
x_{i}-x_{i-1}=\frac{L}{N-1},i=2,...,N\label{fixed}\end{equation}
 Ground states with different interactions (namely, $V$ has a distinguished
minimum and not necessarily nearest neighbors), but without external
force, for classical finite and infinite particle systems were intensively
studied during last 30 years, see for example \cite{Ventevogel_1,Duneau_01},
\cite{Radin_01,Radin_02,Radin_03,Radin_04}. Main results in these
papers concern periodicity of the ground states.

Further on we consider non-zero external force $F(x)\neq0$, and our
main concern is the degree of non-periodicity of the ground states.

\section{Results}

We consider mostly the case when $N$ is sufficiently large. For small
number of particles there can be ugly situations. It is easy to construct
examples with a continuum of fixed points: let $N=3$ and consider
3 particles on the interval $[0,1]$ at the points $x_{1}=0<x_{2}<x_{3}=1$.
Take some $0<y<\frac{1}{2}$ and choose external force as \[
F(x)=f(1-x)-f(x),x\in[y,1-y]\]
 and arbitrary otherwise. Then any $x_{2}$ in $[y,1-y]$ defines
the fixed point $(0,x_{2},1)$.

\begin{lemma}\label{lem1}

For any $N\geq3$ there is not more than one fixed point with $x_{1}=0$
and given $x_{2}$.

\end{lemma}

\begin{lemma}\label{lem2}

There is $N_{0}>0$ such that for $N>N_{0}$ any fixed point $(x_{1},...,x_{N})$
has $x_{1}=0,x_{N}=L$.

\end{lemma}

\begin{theorem}

Assume that $F(x)$ is non-increasing, that is $F(x)\geq F(y)$ if
$x<y$. Then the following assertions hold:
\begin{itemize}
\item fixed point exists and is unique, denote it $(x_{1},...,x_{N})$; 
\item for $k=2,...,N$ as $N\to\infty$ uniformly in $k$\begin{equation}
x_{k}-x_{k-1}\sim\frac{L}{N}\label{quasi_1}\end{equation}

\item if moreover $F(x)$ is smooth and $A>0$ then for any initial conditions
the solution of (\ref{main_eq}) converges to this fixed point; 
\item By $\{x\leftrightarrow L-x\}$-symmetry all these statements hold
for non-decreasing $F$.
\end{itemize}
\end{theorem}

The convergence statement does not concern the main goal of the paper
and is given for completeness.

Only the asymptotics (\ref{quasi_1}) is a bit surprising. The conclusion
is that in the strongly interacting particle system, the macroscopic
external force is not seen on the common microscopic scale (that is
on the scale $\frac{L}{N}$ - the inverse density of particles), but
controlled on much finer scale. One can see this scale from the proof
of the theorem and, more clearly, is exhibited as the second term
of the asymptotic expansion in the next theorem.

\begin{theorem}Assume that $f(x)=x^{-a},$ and that the external
force is constant, that is $F(x)=F$. Then for $k=1,...,N-1$ as $N\to\infty$
\begin{equation}
(x_{k+1}-x_{k})-\frac{L}{N-1}\sim\frac{FL^{a}}{a}N^{-a-1}(\frac{N}{2}-k)\label{quasi_2}\end{equation}

\end{theorem}

\section{Proofs}

\paragraph{Proof of Lemma \ref{lem2}}

Assume $x_{1}>0$. Then the equilibrium condition for the particle
$x_{1}$ is \begin{equation}
f(x_{2}-x_{1})=F(x_{1})\label{fixed_1}\end{equation}
 It follows that $F(x_{1})>0$. By monotonicity of $f$, it follows
from (\ref{fixed_1}) that $x_{2}-x_{1}$ is uniquely defined, and
does not depend on $N$. We also have the following triangular system
- equilibrium conditions for the particles $k=2,...,N-1$

\begin{equation}
f(x_{k+1}-x_{k})=f(x_{k}-x_{k-1})+F(x_{k})\label{fixed_2}\end{equation}
 From (\ref{fixed_2}) we can find by induction $x_{3}-x_{2},...,x_{N}-x_{N-1}$
in a unique way. $F(x_{k})$ can be negative but $f(x_{k}-x_{k-1})+F(x_{k})$
should be non-negative for all $k$, otherwise such fixed point would
not exist. Then for any $k=2,...,N$ we have from (\ref{fixed_1})
and (\ref{fixed_2}) \[
f(x_{k}-x_{k-1})=\sum_{i=1}^{k-1}F(x_{i}),\]
 This gives, for $C=\sup F$, the inequalities \[
0\leq f(x_{k}-x_{k-1})<Ck\Rrightarrow x_{k}-x_{k-1}>(\frac{1}{f(x_{k}-x_{k-1})})^{\frac{1}{a}}>Ck^{-\frac{1}{a}}\]
 If it appears that \begin{equation}
x_{1}+(x_{2}-x_{1})+...+(x_{N}-x_{N-1})=x_{N}=L\label{length}\end{equation}
 then $(x_{1},...,x_{N})$ is a fixed point. If $x_{N}<L$ then\[
F(x_{N})=-f(x_{N}-x_{N-1})\]
 and it is only possible if \[
\sum_{k=1}^{N}F_{k}=0\]
 In any case, $x_{N}\leq L$ is impossible for sufficiently large
$N$, as due to $\frac{1}{a}<1$, the series $\sum k^{-a^{-1}}$ diverges.
For the point $x_{N}$ the proof is symmetric.

\paragraph{Proof of Lemma \ref{lem1}}

Instead of (\ref{fixed_1}) we have for the point $x_{1}=0$ much
weaker condition\begin{equation}
f(x_{2})\geq F(0)\label{boundary}\end{equation}
 Let $x_{2}-x_{1}=x_{2}$ be given such that (otherwise fixed point
could not exist)\[
f(x_{2}-x_{1})+F(x_{2})>0\]
 Then we uniquely find $x_{3}-x_{2}$ from \begin{equation}
f(x_{3}-x_{2})=f(x_{2}-x_{1})+F(x_{2})\label{eq_3}\end{equation}
 Proceeding by induction we find uniquely $x_{3},...,x_{N}$ as functions
of $x_{2}$. If\[
x_{N}<L,f(x_{N}-x_{N-1})+F=0\]
 or\[
x_{N}=L,f(x_{N}-x_{N-1})+F>0\]
 then $(x_{1},...,x_{N})$ is the fixed point, otherwise $(x_{1},...,x_{N})$
is not a fixed point. That is $x_{k},k>2,$ are the functions $x_{k}(x_{2})$
of $x_{2}$.

Now we will prove theorem 1.

\paragraph{Proof of uniqueness}

Put\[
\Delta_{k}=\Delta_{k}^{(1)}=x_{k+1}-x_{k}\]
 We want to prove that the functions $x_{k}(x_{2})$, defined in the
proof of lemma \ref{lem1}, are strictly increasing. Then there is
not more than one value of $x_{2}$ such that $x_{N}=x_{N}(x_{2})=L$,
that gives uniqueness.

For example, if $x_{2}$ increases then the right-hand side of (\ref{eq_3})
decreases, thus $f(x_{3}-x_{2})$ decreases and $\Delta_{2}=x_{3}-x_{2}$
increases. By induction, if $x_{k}<x_{k}^{'},\Delta_{k-1}<\Delta_{k-1}^{'}$
then $F(x_{k}^{'})\leq F(x_{k}),$ $f(\Delta_{k-1})>f(\Delta_{k-1}^{'})$.
It follows from (\ref{fixed_2}) that $f(\Delta_{k})>f(\Delta_{k}^{'})$.
Thus $\Delta_{k}^{'}>\Delta_{k}$ and\[
x_{k+1}^{'}=x_{k}^{'}+\Delta_{k}^{'}>x_{k}+\Delta_{k}=x_{k+1}\]

We know that the function $x_{N}(x_{2})$ is continuous. Thus, to
prove existence one should prove that this function does not grow
too fast with $N$. In this proof we get also the asymptotics for
$\Delta_{k}$.

\paragraph{Existence and asymptotics}

Note that for all $k$ \begin{equation}
f(x_{k+1}-x_{k})=f(x_{2})+\sum_{i=2}^{k}F(x_{i})\label{explicit}\end{equation}
 By (\ref{asympt_f}) for $x\to0$\[
f(x)=\frac{1}{x^{a}}(1+\epsilon(x)),\epsilon(x)=o(1)\]
 We will put for large $N$\[
x_{2}=\frac{L}{N-1}(1+b),b=b(N)=o(1)\]
From (\ref{explicit}) we get that there exists $C>0$ such that for
all $k$ \begin{equation}
-Ck+f(\frac{L}{N-1}(1+b))<f(x_{k+1}-x_{k})<Ck+f(\frac{L}{N-1}(1+b))\label{inequalityMain}\end{equation}
 or \[
(f(\frac{L}{N-1}(1+b))+Ck)^{-\frac{1}{a}}<\Delta_{k}(1+\epsilon(\Delta_{k}))^{-\frac{1}{a}}<(f(\frac{L}{N-1}(1+b))-Ck)^{-\frac{1}{a}}\]
 But\[
(f(\frac{L}{N-1}(1+b))\pm Ck)^{-\frac{1}{a}}=((\frac{L}{N-1}(1+b))^{-a}(1+\epsilon(\frac{L}{N-1}(1+b)))\pm Ck)^{-\frac{1}{a}}=\]
 \[
=\frac{L}{N-1}((1+b)^{-a}(1+\epsilon(\frac{L}{N-1}(1+b)))\pm Ck\frac{L^{a}}{(N-1)^{a}})^{-\frac{1}{a}}=\]
\[
=\frac{L}{N-1}(1+b-\frac{1}{a}\epsilon(\frac{L}{N-1}(1+b))\mp\frac{1}{a}Ck\frac{L^{a}}{(N-1)^{a}}+o(|b|+|\epsilon|+k\frac{L^{a}}{(N-1)^{a}}))\]
 From (\ref{inequalityMain}) it follows that uniformly in $k$ \[
\Delta_{k}=O(\frac{1}{N})\]
Then $\epsilon(\Delta_{k})=o(1)$ also uniformly in $k$, and we have
the following upper bound\[
\sum_{k=1}^{N-1}\Delta_{k}(1+\epsilon(\Delta_{k}))^{-\frac{1}{a}}=\sum_{k=1}^{N-1}\Delta_{k}+o(1)=x_{N}(b)+o(1)\leq\]
\[
\leq L(1+b-\frac{1}{a}\epsilon(\frac{L}{N-1}(1+b)))+\frac{1}{a}C\frac{(N-2)}{2}\frac{L^{a}}{(N-1)^{a}}+o(...))\]
 and similar lower bound with minus in front of $C$. It follows that
the equation \[
x_{N}(b)=L\]
 for $N$ sufficienly large has a unique solution $b$ such that $b=o(1)$.

\paragraph{Proof of Theorem 2}

Here we get exact expression for the second term for the asymptotic
expression of \[
\Delta_{k}=x_{k+1}-x_{k}=\frac{L}{N-1}(1+\delta_{k})\]
 that is the asymptotics for $\delta_{k}$ in case of constant $F$.
We have from (\ref{explicit}) \[
f(x_{k+1}-x_{k})=f(x_{2})+(k-1)F\]
 and instead of (\ref{inequalityMain}) we have ($b=\delta_{1}$)\[
f(\frac{L}{N-1}(1+\delta_{k}))-f(\frac{L}{N-1}(1+b))=(k-1)F\]
 or\[
(1+\delta_{k})^{-a}-(1+b)^{-a}=(\frac{L}{N-1})^{a}(k-1)F\]
 We can rewrite\begin{equation}
1+\delta_{k}=[(1+b)^{-a}+(\frac{L}{N-1})^{a}(k-1)F]^{-\frac{1}{a}}=\label{delta_k}\end{equation}
\[
=[1-ab+O(b^{2})+(\frac{L}{N-1})^{a}(k-1)F]^{-\frac{1}{a}}=1+b-Fa^{-1}(\frac{L}{N-1})^{a}(k-1)+g(b,N,k)\]
where\[
g(b,N,k)=O(b^{2}+b(\frac{L}{N-1})^{a}(k-1)+(\frac{L}{N-1})^{2a}(k-1)^{2})\]
Make summation over $k=1,...,N-1$, and using condition\[
\sum\delta_{k}=0\]
we get\[
\sum_{k=1}^{N-1}\delta_{k}=0=(N-1)b-Fa^{-1}(\frac{L}{N-1})^{a}\sum_{k=1}^{N-1}(k-1)+\sum_{k=1}^{N-1}g(b,N,k)\]
From there\[
b=(N-1)^{-1}Fa^{-1}(\frac{L}{N-1})^{a}\frac{(N-1)(N-2)}{2}+G(b,N)\]
where\[
G(b,N)=O(b^{2}+b(\frac{L}{N-1})^{a}(N-1)+(\frac{L}{N-1})^{2a}(N-1))\]
Thus there is a unique solution for $b$, asymptotically equal to\[
b\sim\frac{FL^{a}}{2a}N^{-a+1}\]
Then from (\ref{delta_k}) we have \[
\delta_{k}\sim\frac{FL^{a}}{a}N^{-a}(\frac{N}{2}-k)\]

\paragraph{Convergence}

Here we will prove that if $A>0$ then for any initial conditions
and $t\to\infty$ the solution converges to the fixed point.

Define the potential energy\[
W(x_{1},...,x_{N})=U(x_{1},...,x_{N})-\sum_{k=1}^{N}\int_{0}^{x_{k}}F(x)dx)\]
Then the equations \begin{equation}
\frac{\partial W}{\partial x_{k}}=0\label{W_minimum}\end{equation}
are the fixed point equations for non boundary points. Note first
that $W$ cannot have minimum in the following cases: 1) $x_{1}>0,x_{N}<L$,
2) $x_{1}=0,x_{N}<L$, 3) $x_{1}>0,x_{N}=L$. This is because the
equations (\ref{W_minimum}), by Lemma 2, do not have solution correspondingly
for 1) $k=1,...,N$, 2) $k=2,...,N$, 3) $k=1,...,N-1$.

It follows that minima can be only if $x_{1}=0;x_{N}=L$. Moreover,
the equations (\ref{W_minimum}) for $k=2,...,N-1$ define both the
fixed point and the minimum of $W$. As these equations have unique
solution, these points coincide.

The known equality

\begin{equation}
\frac{dH}{dt}=\sum_{i}(\frac{\partial H}{\partial p_{i}}\frac{dp_{i}}{dt}+\frac{\partial U}{\partial x_{i}}\frac{dx_{i}}{dt}-F(x_{i})\frac{dx_{i}}{dt})=\sum_{i}(v_{i}\frac{dp_{i}}{dt}+(\frac{\partial U}{\partial x_{i}}-F_{i})v_{i})=\label{Lyap_1}\end{equation}
 \[
=\sum_{i}v_{i}(-\frac{\partial U}{\partial x_{i}}-Av_{i}+F(x_{i}))+\sum_{i}(\frac{\partial U}{\partial x_{i}}-F(x_{i}))v_{i}=-A\sum v_{i}^{2}\]
 shows that the full energy $H=\frac{p^{2}}{2m}+W$ is non-increasing.

Define metrics in the set $\{X=(x_{1},...,x_{N})\}$ of configurations\[
\rho(X,Y)=\sum_{k=1}^{N}|x_{k}-y_{k}|\]
 We want to show that $H$ reaches any neighborhood of the minimum
for finite time.

\begin{lemma}

For any $\epsilon>0$ there are constants $0<\delta=\delta(\epsilon),\tau=\tau(\epsilon)<1$
such that for any initial conditions outside the neighbourhood $O(\epsilon)$
of the fixed point \[
\int_{0}^{\tau}\sum_{k=1}^{N}v_{k}^{2}(t)dt>\delta\]
 \end{lemma}

From this lemma the last assertion of Theorem 1 evidently follows.

Proof of the Lemma. For given $X$ denote $F_{k}=F(x_{k-1},x_{k},x_{k+1})$
the force acting on the particle $k$ at the point $x_{k}$. Denote\[
\Phi_{0}=\Phi_{0}(X)=\sum_{k=2}^{N-1}|F_{k}|,\Phi_{1}(X_{9})=\sum_{k=2}^{N-1}|\frac{\partial F_{k}}{\partial x_{k}}|\]
 and consider the set of configurations\[
A=A(D_{1},D_{2})=\{X:\Phi_{0}<D_{0,},\Phi_{1}<D_{1}\}\]
 Fix some $\epsilon>0$ sufficiently small. Then, by smoothness, there
exists constant $0<c_{1}=c_{1}(\epsilon)<\infty$ (depending on $N$)
such that for any $X\in A\setminus O(\epsilon)$ \begin{equation}
\Phi_{0}>c_{1}\label{F_bounds}\end{equation}
 Take some initial conditions $(x_{1},v_{1}),...,(x_{N},v_{N})$ with
$X=(x_{1},...,x_{N})\in A\setminus O(\epsilon)$ and let $k$ be such
that $F_{k}=F_{k}(X)$ is maximal among other $F_{j}$, then \[
|F_{k}|>\frac{c_{1}}{N}\]
 The idea is quite simple: if non of the points $y_{k-1},y_{k+1}$
does not get sufficient speed, then the force $F_{k}$ will give sufficient
speed to the particle $k$. More exactly, for \[
\delta_{1}=\frac{F_{k}}{2D_{2}}\]
 we have\[
|F_{k}(y_{k-1},y_{k},y_{k+1})|>\frac{1}{2}|F_{k}(x_{k-1},x_{k},x_{k+1})|\]
 for any $y_{i},i=k-1,k,k+1$ with $|x_{i}-y_{i}|<\delta_{1}$. Then
if none of the three points $y_{k-1},y_{k},y_{k+1}$ on the time interval
$(0,\tau)$ passes the distance greater than $\delta_{1}$, then $y_{k}$
will pass the distance not smaller than \[
\delta_{1}=\frac{F_{k}\tau^{2}}{2}\]
 Put then\[
\tau=\sqrt{\frac{2\delta_{1}}{F_{k}}}\]
 Then we have the alternative: either \[
\int_{0}^{\tau}|v_{k}|d\tau>\delta_{1}\]
 or\[
\int_{0}^{\tau}(|v_{k-1}|+|v_{k+1}|)d\tau>\delta_{1}\]
 Thus \[
\int_{0}^{\tau}(|v_{k-1}|+|v_{k}|+|v_{k+1}|)dt>\delta_{1}\]
 Now one can use the Cauchy inequality\[
\int_{0}^{\tau}v^{2}d\tau\geq\frac{1}{\tau}(\int_{0}^{\tau}|v|d\tau)^{2}\]
 We remark that no reasonable convergence rate follows from this proof.

\section{Some remarks}

The appearence of finer scale (we call it submicroscale) with macroeffects
is now a rare phenomenon. In many equilibrium models of infinite particle
systems local equilibrium conditions are defined on the normal microscale.
For example, the corresponding problem (for ground states and for
Gibbs low temperature states) in the one-dimensional model of elasticity
(the Hooke's law), discussed in \cite{Mal_Hooke} where the force,
acting only on the particle $x_{N}$, produces discrete laplacians\[
\Delta_{k}^{(2)}=\delta_{k+1}-\delta_{k}=x_{k+1}-2x_{k}+x_{k-1}\]
of the order $\frac{1}{N}$.


\begin{thebibliography}{8}
\bibitem{Mal_Current}V. A. Malyshev. Why current flows: a multiparticle
one-dimensional model. Theoretical and Mathematical Physics, 2008,
v. 155, No. 2, 766-774.

\bibitem{Mal_Hooke}V. A. Malyshev. One-dimensional mechanical networks
and crystals. Moscow Math. Journal, 2006, v. 6, No. 2, pp. 353-358.

\bibitem{Ventevogel_1}W. Ventevogel. On the configuration of a one-dimensional
system of interacting particles with minimum potential energy per
particle. Physica A, 1978, 92, No. 3-4, pp. 343-361.- 

\bibitem{Duneau_01}M. Duneau, A. Katz. Structural stability of classical
lattices in one-dimension. Annales de l'I.H.P., section A, 1984, 41,
No. 3, 269-290.

\bibitem{Radin_01}Ch. Radin. Existence of ground state configurations.
Math. Physics Electronic J., 2004, v. 10.

\bibitem{Radin_02}Ch. Radin. Crystals and Quasicristals: a lattice
gas model. Physics Letters, 1986, 114A, No. 7, 381-383.

\bibitem{Radin_03}C. Gardner, Ch. Radin. The infinite-volume ground
state of the Lennard-Jones potential. J. Stat. Phys., 1979, v. 20,
No. 6, 719-724.

\bibitem{Radin_04}Ch. Radin, L. Schulman. Periodicity of classical
ground states. Phys. Rev. Letters, 1983, 51, No. 8, 621-622. 
\end{thebibliography}
\end{document}